\documentclass[twocolumn,secnumarabic,amssymb, nobibnotes, aps, prd]{revtex4-2}

\newcommand{\env}[1]{\texttt{#1}}
\usepackage{graphicx}
\usepackage{dcolumn}
\usepackage{bm}%
\usepackage{amsmath}
\usepackage[caption = false]{subfig}
\usepackage[utf8x]{inputenc}
\usepackage{xcolor}
\usepackage{amssymb}
\usepackage{natbib}
\begin{document}

\preprint{APS/123-QED}

\title{Effects of Nontrivial Topology on Neutron Star Rotation and its Potential Observational Implications}

\author{Debojoti Kuzur}
 \email{dkuzur.phys@raghunathpurcollege.ac.in}
\affiliation{Department of Physics, Raghunathpur College, Purulia, West Bengal 723133, India.
}%
\date{\today}

\begin{abstract}
Rotational irregularities are one of the prominent observational features that most pulsars exhibit. These glitches, which are sudden increases in spin angular velocity, remains an open problem. In this study, we have investigated the potential role of nontrivial topological defects, specifically in the form of Nambu-goto-type CSs, and its connection to spin irregularities. Such CSs which are one-dimensional topological defects may be formed during various symmetry-breaking and phase transition scenarios and can interact with the neutron stars.
In this work, we see that the appearance of such topological defects trapped within the core can lead to the coupling of the string tension with the angular velocity, leading to the abrupt rotational changes observed as pulsar glitches.
We have further studied how these coupling may generate detectable gravitational waves as a mixture of continuous and burst signals. The evolution of cusps of CSs trapped within neutron stars and the neutron star's mass quadruple moment change due to rotation could produce distinctive gravitational wave signatures, well within the noise cutoff of advLIGO. Our study highlights a potential connection between topological defects, pulsar glitches, and gravitational wave emissions, offering a possible avenue for observationally testing the presence of CSs and their astrophysical effects.
\end{abstract}

\maketitle

\section{Introduction}
Compact stars such as Neutron stars (NS) formed from core collapsing supernova explosions have highly dense cores. Such NS has extreme physical conditions, like exceeding nuclear saturation densities, around $10^{15}$ g/cm$^3$ or 1 baryon/fm$^3$ number density, strong gravitational fields, and high rotational speeds, making them natural laboratories for testing the limits of fundamental physics \cite{Lasky_2015,annurev:/content/journals/10.1146/annurev-nucl-102419-124827,annurev:/content/journals/10.1146/annurev-nucl-102711-094901,annurev:/content/journals/10.1146/annurev-astro-081915-023322,annurev:/content/journals/10.1146/annurev.aa.04.090166.002141}. Under such extreme condition, one of the model for the core is the overlapping of the baryons and pointing to a state of de-confined quark matter (QM) resulting in a phase transition (PT) \cite{PhysRevC.107.025801,doi:10.1142/S0218301318300084, PhysRevD.107.083023, Orsaria_2019, GLENDENNING2001393, sym16010111, MALLICK201496}. Various constraints on the maximum mass and radius of such stars has been implemented \cite{PODDER2024122796,Pan_2007} from the gravitational wave data of the binary merger GW 190814 \cite{Tangphati2022}, and from the observed pulsars PSR J0740+6620, PSR J0952-0607, and PSR J0030+0451 using both Bayesian analysis \cite{PhysRevD.109.043054} and also using the frame work of Einstein–Gauss–Bonnet gravity \cite{10.1093/mnras/stac3611}. 

Recent studies in the field of high-energy astrophysics and condensed matter physics have shown the possible presence and significance of topological phases in strongly interacting matter, such as those found in the NS cores \cite{condmat7010026,PhysRevB.109.064503,10.1093/nsr/nwt033,PhysRevB.99.125122}. The presence of stable, global features which cannot be removed by continuous deformations, describes a nontrivial topology such as vortex lines, domain walls, or cosmic strings (CS). In the case of NS, such topological defects (TD) are expected to form in the core, where densities are high enough to form color superconducting phases or QM and are governed by the principles of quantum chromodynamics (QCD) \cite{ZHITNITSKY1999647,doi:10.1142/9789812701725_0005,Higaki2016}. 

In this work we have considered CSs which are hypothetical one-dimensional TDs that may have formed during PTs in the early universe. These defects, which are a consequence of symmetry-breaking processes, are modeled using the Nambu-Goto (NG) action expressed as the area of the string’s world-sheet, a two-dimensional surface traced out by the string as it moves \cite{Vachaspati:2015,Blanco-Pillado_2023,Sanyal2022,PhysRevD.91.083519,zwiebach2009first}. The action is written as:
\begin{equation}
    S=-\mu\int d^2\sigma\sqrt{-g}
\end{equation}
where, $\mu$ is the energy per unit length of the string, $\sigma$ is the coordinates of the two dimensional world-sheet traced by the string and $g$ is the determinant of the metric induced on the world-sheet.
Transitions from a high-energy symmetric state to a lower-energy state with broken symmetry, described mathematically by the breaking of a larger symmetry group $G$ to a smaller subgroup $H$. Now for a scalar field $\Phi$ whose dynamics are governed by a potential with spontaneous symmetry breaking, at high energies, the field respects the symmetry group $G$. As the universe cools, the potential causes the field to settle into a vacuum manifold $M=G/H$, where $H$ is the unbroken subgroup. CSs are associated with non-contractible loops in this vacuum manifold, which can be classified by the fundamental group $\pi_1(M)$. If $\pi_1(M)\ne 0$, then the vacuum manifold supports CS solutions. Typically the breaking of a $U(1)$ symmetry, where the vacuum manifold $M$ is a circle $S^1$, whose fundamental group is the non-trivial, $\pi_1(S^1)=\mathbb{Z}$. This non-trivial fundamental group implies that strings can form, and their winding number corresponds to elements of the group $\mathbb{Z}$, representing how many times the field winds around the vacuum manifold \cite{PhysRevLett.66.1126,WITTEN1985243,SARANGI2002185,VILENKIN1985263,Achúcarro2009,Hindmarsh_1995}.

Accordingly such TDs if formed in these phases arising from PT via the Kibble-Zurek mechanism in the core of NS \cite{Dave_2019,PhysRevX.5.021015,10.1093/mnras/stae1642}, could significantly affect the star’s internal structure and thus its rotational dynamics. NS rotate rapidly, often with periods ranging from milliseconds to seconds, and exhibit phenomena such as ``glitches" which are sudden changes in their rotational speed. These glitches are thought to result from angular momentum exchange between different components of the star, such as the crust and the superfluid interior \cite{Marmorini2024,doi:10.1142/S0218271815300086,HASKELL2024102921,Antonopoulou2022-vs,universe8120641}. Recent theoretical work suggests that the presence of non-trivial topology within NS could modify their moment of inertia, leading to changes in their rotational frequency. For example, vortex lines in a superfluid phase may act as conduits for angular momentum transfer, while domain walls or other TDs could interact with the star’s magnetic field, leading to the dissipation or storage of rotational energy. These effects could manifest as changes in the star’s spin-down rate, or they may influence the timing and magnitude of observed glitches \cite{PhysRevLett.116.169002,Srivastava2017,10.1093/mnras/sty130,Pizzochero_2011}. Moreover, TDs could also affect the coupling between different layers of the star, further impacting its rotational dynamics. The potential observational implications of non-trivial topology in NS are quite possible. One of the most promising signatures of topological effects is the emission of gravitational waves. If TDs exist in the core, they could interact with the star’s rotation and magnetic field, producing gravitational wave signals that could be detectable by observatories like advLIGO and Virgo \cite{Chang2022,Gouttenoire2022,Auclair_2023,PhysRevD.110.063549,Sousa2024,PhysRevLett.85.3761}. 

In summary, the study of NS glitches and spin-down irregularities could provide indirect evidence for the presence of topological structures and non-trivial topology. By linking topological phases of matter with observable astrophysical phenomena, some new insights could be gained into the physics of dense matter, the properties of quantum fields under extreme conditions, and the broader implications for astrophysical and cosmological models. The paper is structured as follows: In section II, we have defined the metric of a compact object coupled to a NG type defect. We have solved such a metric to obtain the rotational equation for the star and have also discussed various EoS that can arise due to a hadronic and a hybrid star. In section III, we have solved the rotational equation for various EoS and also across various energy density for the strings. Here we have also discussed about the universality of the rotational changes across EoSs. In section IV, we have discussed how the appearance of a TD can cause the spin up of the star and how radiative losses due to such strings can cause the star to lose angular momentum. In section V, we have discussed the possible gravitational wave signals for such a star and the detectibility of it in the advLIGO and Virgo detectors. In section VI we have drawn conclusions from the results obtained.

\section{Spacetime Framework}
\subsection{Metric}
We start by considering the metric for the NG type defect which is typically expressed in cylindrical coordinates $(t,r,\theta,z)$. Converting the metric from cylindrical coordinate system into a spherical one $(t,r,\theta,\phi)$, it is then used to form the background of a spherically symmetric, rotating NS. The metric for such a system can be written (using geometrized units for the rest of the paper, by taking $G=c=1$) as,
\begin{equation}
    ds^2=-e^{2\Phi}dt^2+e^{2\Lambda}dr^2+r^2d\theta^2+r^2\sin^2\theta[\alpha d\phi-\omega dt]^2
\end{equation}
where $\Phi=\Phi(r)$ and $\Lambda=\Lambda(r)$ are the metric potentials, and $\omega=\omega(r)$ is the frame dragging term coupled with the time coordinate of the metric. This metric has the usual form of an axisymmetric spacetime for a rotating compact object. However, the non-trivial aspect here arises from the background NG defect reflected, by the term $\alpha=(1-4\mu)$, coupled to the $\phi$ coordinate, where $\mu$ is the energy density of the string. This kind of defect creates a topology such that the coordinate $\phi$ has the range $0<\phi<2\pi(1-4\mu)$.

For a slowly rotating NS, the metric can be expanded upto $O(\omega^2)$ and can be written as,
\begin{equation}
    g_{\mu\nu}=\begin{pmatrix}
-e^{2\Phi} & 0 & 0 & -\alpha\omega r^2\sin^2\theta\\
0 & e^{2\Lambda} & 0 & 0\\
0 & 0 & r^2 & 0\\
-\alpha\omega r^2\sin^2\theta & 0 & 0 & \alpha^2r^2\sin^2\theta\\
\end{pmatrix}
\end{equation}
The Christoffel symbols, Ricci Tensor, Ricci Scalar and the Einstein Tensor are calculated, with the non-triviality in terms of $\alpha$ only appearing in the $G_\phi^t$ component upto $O(\omega^2)$ 
\subsection{Stress Energy Tensor}
For the right hand side of the Einstein's field equations, we calculate the stress energy tensor $T_{\mu\nu}$ for the NS coupled to a NG defect. The NS is approximated as a perfect fluid and the stress energy is taken to be
\begin{equation}
    T_{\mu\nu}^{NS}=(\rho+P)u_\mu u_\nu + Pg_{\mu\nu}
\end{equation}
where $\rho=\rho(r)$ and $P=P(r)$ are the density and pressure of the fluid. The $u_\mu$ are the four-velocities of the fluid components. The velocities are defined as $u_\mu=\frac{dx^\mu}{d\tau}$ with respect to the proper time $\tau$ and can b expanded as, $u_\mu=\frac{dx^\mu}{dt}\frac{dt}{d\tau}$. Due to the axisymmetry nature, the fluid components of the star doesn't have a motion along the $r$ and $\theta$ directions, thus $u_r=u_\theta=0$. In the direction of the symmetry $\phi$, $u_\phi=\Omega u_t$ where, $\frac{d\phi}{dt}=\Omega$, is the angular velocity of the star. 

The stress energy tensor for the background defect can be written as \cite{VILENKIN1985263} 
\begin{equation}
    T^{DE}_{\mu\nu}=\mu \delta(r)\times diag(1,0,0,1)
\end{equation}
where $r=0$ is the position of the string arising from the TD. In spherical coordinate system upto $O(\omega^2)$,
\begin{equation}
    T^{DE}_{\mu\nu}=\begin{pmatrix}
\mu & 0 & 0 & 0\\
0 & \mu\cos^2\theta & -r\mu\cos\theta\sin\theta & 0\\
0 & -r\mu\cos\theta\sin\theta & r^2\mu\sin^2\theta & 0\\
0 & 0 & 0 & 0\\
\end{pmatrix}
\end{equation}
Thus the total stress energy tensor of the system is
\begin{equation}
    T_{\mu\nu}=T^{NS}_{\mu\nu}+T^{DE}_{\mu\nu}
\end{equation}
This is the stress energy tensor for a rotating NS with a TD arising from the core of the star. It can be seen that $00$-component of the $T^{DE}$ corresponds to the energy density of the string $\mu$ and the absence of $\phi\phi$-component of the tensor reflects the axi-symmetry of the system.
\subsection{Equation for rotation}
The rotational equation is obtained from the $t\phi$ component of the Einstein's Field equation
\begin{equation}
    G^t_\phi=8\pi T^t_\phi
\end{equation}
which gives the following second order differential equation,
\begin{equation}
    \frac{d^2\omega(r)}{dr^2}+D_1\frac{d\omega(r)}{dr}-D_2(\Bar{\Omega}+\omega(r))=0
    \label{omega}
\end{equation}
where,
\begin{align}
    &D_1=\frac{4-rB(r)}{r}\\
    &D_2=\frac{16\pi r A(r)e^{2\Lambda}}{r}\\
    &\Bar{\Omega}=(4\mu-1)\Omega
    \label{omgbar}
\end{align}
Inside the star, that is $0< r\le R$, where $R$ is the radius of the star, the functions $A$ and $B$ are defined as $A(r)=P(r)+\rho(r)$ and $B(r)=\Lambda'(r)+\Phi'(r)$ and the metric potentials are $\Lambda=-\frac{1}{2}ln\left[1-\frac{2m(r)}{r}\right]$ and $\Phi'(r)=-\frac{1}{A(r)}P'(r)$. The coupling of the defect with the rotation can be seen from equation \ref{omgbar}. This equation is solved numerically along with the TOV equations for pressure and density till the radius of the star is obtained. The surface of the star is identified with $P(r)\rightarrow 0$ and $\rho(r)\rightarrow 0$. For $r\ge R$, the term $D_2\rightarrow 0$ and hence the known analytical solution of equation \ref{omega} is obtained, $\Omega=\omega+\frac{R}{3}\frac{d\omega}{dr}\Big|_R$. The $\omega(R)$ has to be continuous on surface of the star and thus the value of $\omega(r)$ at the center is adjusted in such a way to get a desired $\Omega$.
\subsection{Neutron Star Equation of state}
In order to numerically solve equation \ref{omega} along with the TOV equations, a class of EoS has been taken respecting the current observational constraints. 
\begin{figure}[h!]
    \centering
    \includegraphics[scale=0.6]{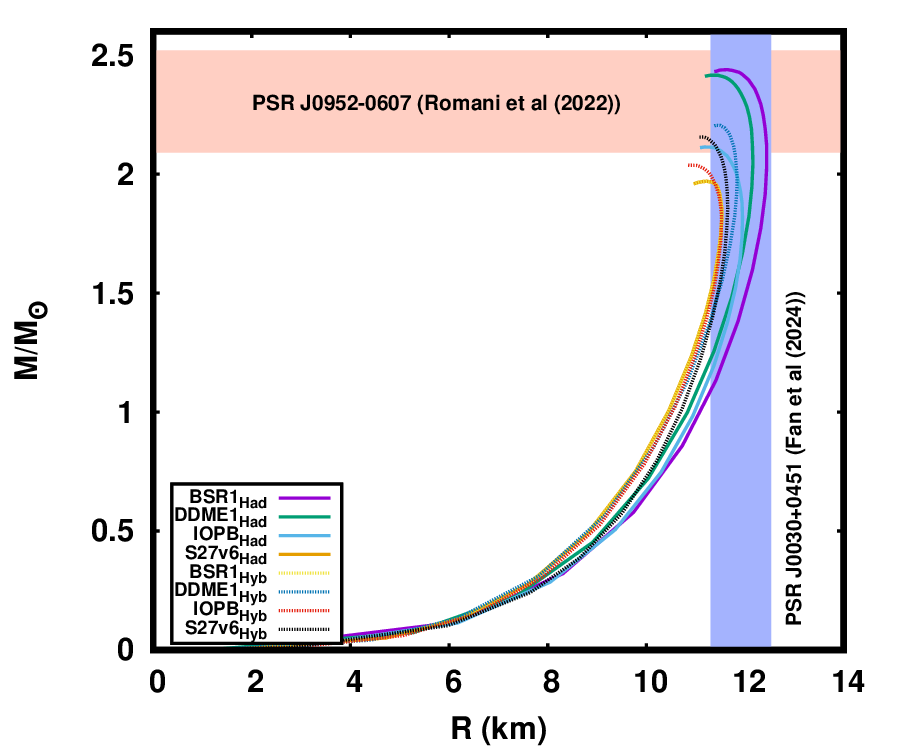}
    \caption{The curves represent different theoretical models of the (EOS) of NS matter. These models include both hadronic (Had) and hybrid (Hyb) EOS. The shaded region at the top around 2.35 solar masses corresponds to the mass measurement of PSR J0952-0607, which is currently considered one of the heaviest known NSs. This region serves as an upper constraint on the NS mass. The blue vertical band around 12 km corresponds to the radius constraint derived from observations of PSR J0030+0451, based on NICER measurements. This provides a constraint on the NS radius. Almost all curves fall within the observed mass-radius limits suggests that these models remain viable under current observations.}
    \label{fig1}
\end{figure}
The hadronic BSR equation of state \cite{PhysRevC.76.045801} is a field theoretical based relativistic mean field model which is taken for $\beta$ - equilibriated nucleons and hyperons in the high density regime. The hadronic DDME EoS \cite{PhysRevC.66.064302} is based on the relativistic random phase approximation which incorporates a density dependent meson nucleon vertex function. In the Hadronic IOPB EoS \cite{PhysRevC.97.045806}, simulated annealing method is used to create a parameter set that predicts the general properties of nuclear matter under the Effective mean field theory. The S27 hadronic EoS \cite{PhysRevC.66.055803} on the other hand obtains a symmetry energy correlated with the accuracy in measurement of the neutron radius of ${}^{208}$Pb nucleus. In all of the EoS, at low densities we have added the BPS EoS \cite{1971ApJ...170..299B} in order to define the crust. The QM EoS is approximated by the MIT bag model, which is constructed in such a way that the hadronic matter, mixed phase and pure QM are present in the low, medium and high densities respectively. The QM EoS is parameterized by the effective bag constant $B_{eff}$ and constant $a_4$ which incorporates the strong interaction and non-perturbative QCD effects.

Varying the central densities, we have plotted the M-R curve (figure \ref{fig1}) for all the classes of EoS as mentioned above. In the figure we have also shown the current constraints for the radius and mass of the NS. The maximum mass data of NS has been reported by the fastest rotating pulsar PSR J0952-0607 (in the galaxy) \cite{Romani_2022} having a mass $M=2.35 \pm 0.17 M_{\odot}$ whereas the radius constraint is reported to be $R=11.90^{+0.63}_{-0.60}$ km \cite{PhysRevD.109.043052}. In this work all of the EoS (Hadronic and Hybrid) satisfies the observational constraints as discussed. The constraints obtained from PSR J0740+6620 by the NICER observations \cite{Riley_2021,Miller_2021} are also in agreement with the choice of the EoS, where as the constraints setup by the gravitational wave event GW190814 \cite{Abbott_2020} is satisfied only by couple of hadronic EoS although the secondary component has yet to be identified as a NS or a BH.
\section{Coupling of rotation with the defect}
\begin{figure*}[ht!]
    \includegraphics[scale=0.54]{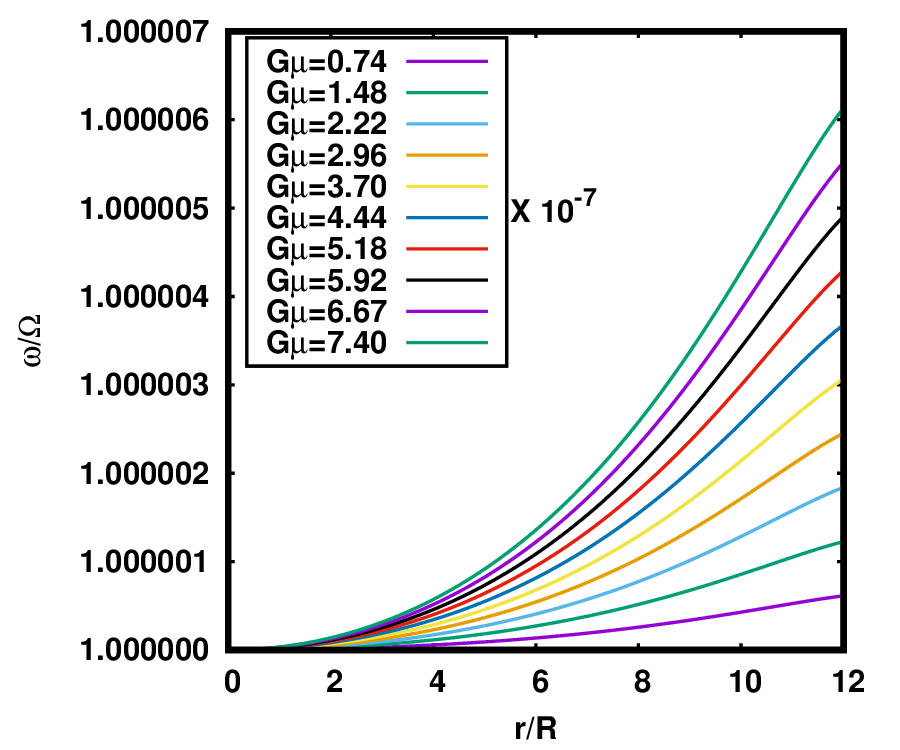}
    \hspace{1 cm}
    \includegraphics[height=6.5cm,width=8cm]{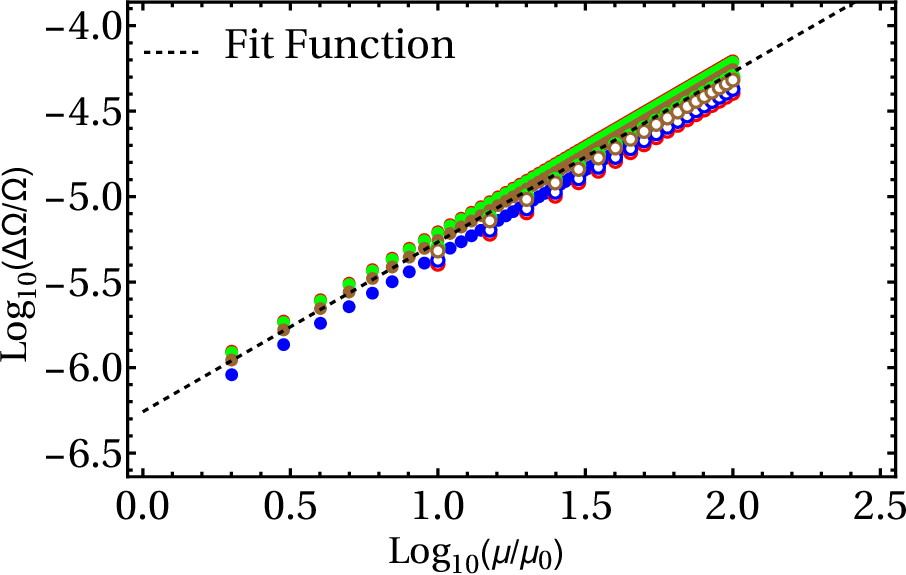}
    \caption{(left panel) The plot represents a model of rotational frame dragging for a compact star normalized with the central angular velocity, with varying values of $G\mu$ (order of $10^{-7}$) corresponding to different levels of gravitational tension or the energy density of the string. As the radial distance from the center increases, the frame dragging velocity slightly exceeds the central angular velocity, and this effect becomes more pronounced as $G\mu$ increases implying that in systems with stronger gravitational interactions (or higher energy density), rotational effects are more pronounced, particularly in the outer regions. (right panel) The plot represents the relationship between the fractional change in angular velocity $\Delta\Omega/\Omega$ and a normalized energy density of the string tension $\mu/\mu_0$. This plot demonstrates a power-law relationship due to the linear trend on the log-log scale and the fit function models an increasing power law. The consistency of the data points across various EoSs suggests that this relationship holds over a wide range of conditions, making it a robust characteristic.}
    \label{fig2}    
\end{figure*}

The class of EoS along with equation \ref{omega} is then numerically solved till the surface of the star R. For a given choice of initial angular velocity $\Omega=\Omega_{int}$, the frame dragging effect $\omega(r)$ of the NS has been shown in (figure \ref{fig2} (left panel)). The variation of $\omega(r)$ can be seen as the value of $G\mu$ increases. Here although till now for ease of calculations we were working in a $G=c=1$ unit system, however traditionally, the discussion of TDs involves the dimensionless quantity $G\mu$ and thus from now on will restrict our discussions only to $c=1$ unit system.     

The value of $G\mu$ shown here is restricted to the constraints imposed by the recent CMB observations, where the bound is $G\mu < 6.1\times 10^{-7}$ \cite{Levon_2009,Jeong_2005}. The frame dragging $\omega(r)$ is normalized with the initial value of $\Omega$ and the ratio $\omega(r)/\Omega$ can be seen increasing with the increase in $G\mu$. This coupling of the defect with the rotation can be seen from the term $(4\mu-1)\Omega$ in equation \ref{omgbar} which in the limit $\mu\rightarrow 0$ boils down to the usual frame dragging equation \cite{1967ApJ...150.1005H}. The rotational differential equation \ref{omega} can be compared to have the general from $a\Ddot{X}+b\Dot{X}+cX=0$ where the second term depicts the usual damping or amplifying effect and the third term acts as a restoring influence which tires to bring the system to its original configuration. The defect in form of $\mu$ enters this restoring term where if $\mu=0$, we have $c>0$ and the third term acts to restore the system to its initial state and thus deviates less from its initial value of $\omega(r=0)$. However due to the presence of the defect, for values $4\mu-1>0$ (which is mostly the case in our work), the coefficient $c<0$ and the restoration effect reduces and thus a significant deviation can be seen for $\omega(r)$.
The final angular velocity can be obtained from the frame dragging velocity as $\Omega=\omega+\frac{R}{3}\frac{d\omega}{dr}\Big|_R$. The change in the angular velocity $\Delta \Omega$ from its initial value (at $G\mu=0$), normalized by the initial angular velocity $\Omega$ is shown (figure \ref{fig2} (right panel)). It is plotted with respect to increasing $G\mu$ normalized with a standard value $G\mu_0$. The relation is almost linear with small variation across various class of EoS (both hadronic and hybrid) in the log-log scale. The semi-universality of this relation shows a strong co-relation between the change in angular velocity and $G\mu$. The fit function obtained for the plot is of the form
\begin{equation}
    y=k_1 x + k_2
\end{equation}
where $k_1=0.993$ and $k_2=-6.258$ with the reduced chi squared value of $\chi_{red}^2=6.4\times 10^{-5}$. From this fit data, energy density or the gravitational strength of the defect within the core of the NS can be obtained from the change in angular velocity as
\begin{equation}
    G\mu=10^{\left[\frac{log_{10}(\Delta\Omega/\Omega)+6.258}{0.993}\right]}G\mu_0
    \label{gmu}
\end{equation}
A power law relation was expected due to the linearity in the log-log scale. In the next section, this relationship is used to obtain various estimates of the $G\mu$ that is physically possible from the parameters obtained from variety of observed pulsars. 
\section{Spin-up and radiative losses}
A TD when identified as a CS, can lead to some radiative processes. If the NS has a dipolar magnetic field associated with it, then the rotating NS would lead to dipole radiation loss which is given by \cite{PhysRevD.91.063007},
\begin{equation}
    \frac{dE}{dt}\Big|_{dip}=-\frac{2}{3}a^2\Omega^4\sin^2\alpha
\end{equation}
where $a$, and $\alpha$ are the magnetic dipole moment and the angle of inclination between rotation axis and dipole moment respectively. This energy loss in general causes the magnetic breaking thus monotonically decreases $\Omega$ over time. Now the appearance of the TD during PT leading to the CS can further produce additional energy losses in terms of gravitational wave by oscillation of the string loops given as \cite{PhysRevD.71.063510}
\begin{equation}
    \frac{dE}{dt}\Big|_{def}=-\Gamma G\mu^2
\end{equation}
where $\Gamma$ is a numerical constant. The total energy loss due to dipolar magnetic field and string oscillation has combined effect on the angular velocity $\Omega$. In order to quantify the effects we write the total energy loss as:
\begin{equation}
    \frac{dE}{dt}\Big|_{tot}=\frac{dE}{dt}\Big|_{dip}+\frac{dE}{dt}\Big|_{def}
\end{equation}
\begin{figure*}[ht!]
    \centering
    \includegraphics[scale=0.53]{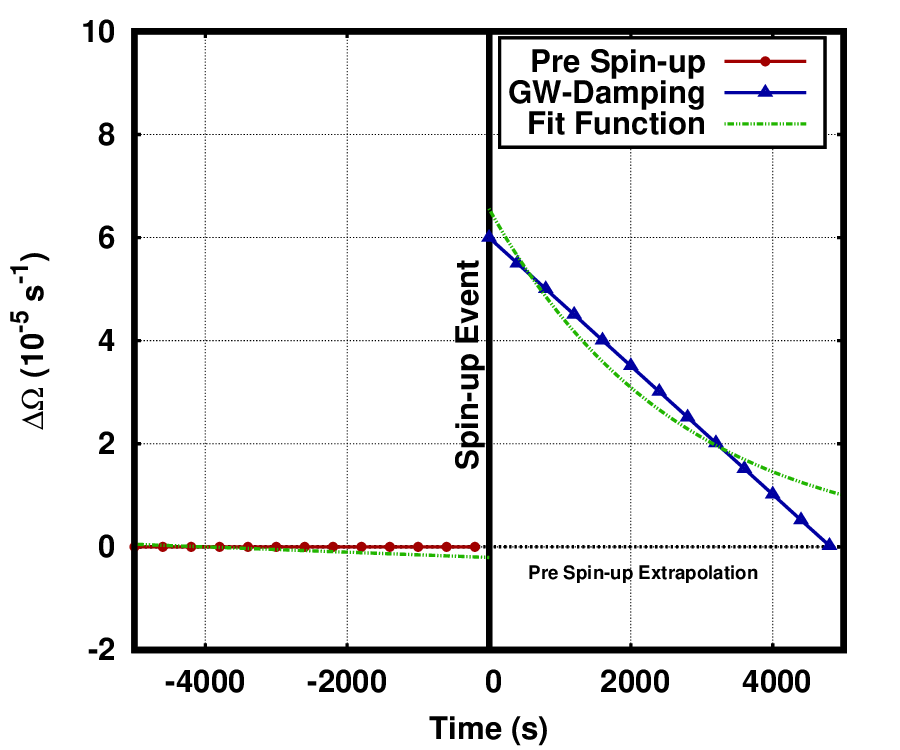}
    \hspace{1 cm}
    \includegraphics[scale=0.53]{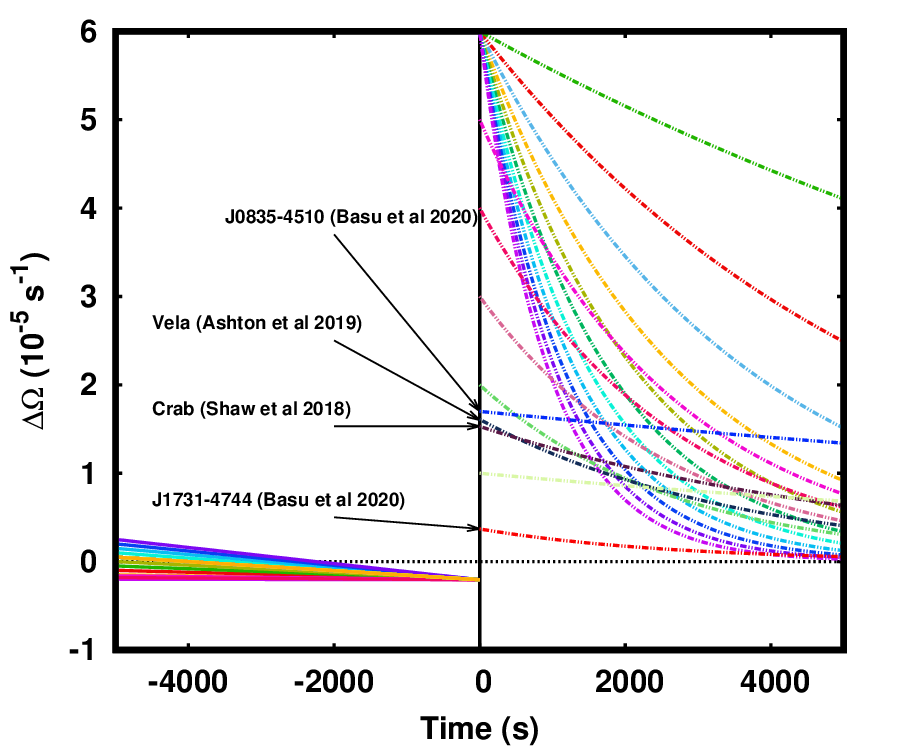}
    \caption{(left panel) The plot illustrates a time-based analysis of the change in the angular velocity of the NS $\Delta\Omega$ measured in $10^{-5}$ s$^{-1}$. The plot consists of two distinct time regimes: pre-spin-up and post-spin-up events. At $t=0$ s, a significant change in $\Delta\Omega$ is initiated due to the appearance of a TD, marking the onset of a spin-up event. The spin-up event induce a perturbation or decay behavior in $\Delta\Omega$ indicating the influence of a damping mechanism due to the gravitational wave damping from CS cusps, which affects the system’s angular velocity. The fit function used effectively models this decay, and can be used as a predictive tool for analyzing similar phenomena. (right panel) The plot shows analysis of changes in $\Delta\Omega$ as a function of time, for family of curves each with different angular velocity change and damping rate (corresponding to particular $G\mu$ and gravitational radiation rate). Several pulsars are overlapped and labeled, with curves representing their behaviors during and after a "glitch" event. The very slow decay in pre-spin-up indicates a gradual energy loss, due to dipole radiation.}
    \label{fig3}
\end{figure*}
Now the total energy for a rotating star in terms of angular velocity $\Omega$ and moment of inertia $I$ can be written as $E=1/2 I\Omega^2$, which gives,
\begin{equation}
    \frac{d}{dt}\left(\frac{1}{2}I\Omega^2\right)=-\frac{2}{3}a^2\Omega^4\sin^2\alpha-\Gamma G\mu^2
\end{equation}
Simplifying this in order to obtain a differential equation for $\Omega$, we can write
\begin{equation}
    \frac{d\Omega}{dt}=-\frac{2}{3}\frac{a^2}{I}\Omega^3\sin^2\alpha-\Gamma \frac{G\mu^2}{I\Omega}
\end{equation}
In order to rewrite the entire right hand side in terms of $\Omega$, we have to use some scaling factors. Now the gravitational wave energy can be scaled as $E\sim \mu l$, where $l$ is the typical size of the string loops that are formed \cite{PhysRevD.71.063510}. From here we find that the energy density $\mu\sim \frac{1}{2}\frac{I\Omega^2}{l}$ and thus the differential equation for angular velocity can be written as
\begin{equation}
    \frac{d\Omega}{dt}=-(A+B\Theta(t-t_0))\Omega^3
    \label{domeg}
\end{equation}
where, $\Theta(t-t_0)$ is the Heaviside step function and $t_0$ is the time at which the string loop appears and,
\begin{align}
    &A=\frac{2}{3}\frac{a^2}{I}\Omega^3\sin^2\alpha\\
    &B=\frac{1}{4}\frac{\Gamma G I}{l^2}
\end{align}
From the equation \ref{domeg} it is seen that during the PT, as the defect and hence the string appears in the core of the NS, the angular velocity of the NS $\Delta\Omega$ changes and we will obtain a spin up event, followed by the release of gravitational wave at $t>t_0$. This in turn tries to bring back the angular velocity $\Omega$ back to $\Omega_{int}$ because the loss in the form of GW energy (also dipole radiation energy) is compensated from rotational energy. For the entire series of event, that is, pre-spinup, during spinup and GW damping, we have fitted a functional form as shown in (figure \ref{fig3} (left panel)). The fit function thus obtained is,
\begin{equation}
f(t)=\begin{cases} 
      a_1t+b_1 & t\le t_0 \\
      a_2e^{-b_2(t-t_0)}+c_2 & t > t_0
   \end{cases}
\end{equation}
with the values, $a_1=-5.12\times 10^{-5}$, $b_1=-0.205$, $a_2=6.56$ and $b_2=3.76\times 10^{-4}$ with a reduced chi squared value of $\chi^2_{red}=2.2\times 10^{-5}$. Using this fit function and the set of parameters $(a_1,b_1,a_2,b_2)$, we have plotted a (non-exhaustive) family of curves with varying combinations of the parameter set (figure \ref{fig3} (right panel)) out of which we have labeled four pulsars PSR J0835-4510 \cite{10.1093/mnras/stz3230}, PSR B0833-45 (Vela Pulsar) \cite{Ashton2019}, PSR B0531+21 (Crab Pulsar) \cite{10.1093/mnras/sty1294} and PSR J1731-4744 \cite{10.1093/mnras/stz3230} whose glitch profile is similar to that particular curve in the family. Comparing the fit function with the standard model for pulsar timing technique for a glitch event \cite{universe8120641}
\begin{equation}
    \phi_g(t)=\Delta\phi_g+\Delta\Omega_p(t-t_0)-\Big[\Delta\Omega_d\tau_d\left(1-e^{-(t-t_0)/\tau_d}\right)\Big]
\end{equation}
where, $\phi_g$ is the rotational phase, $\Delta\Omega_p$ is the jump in the spin frequency occurring at time $t_0$ and $\Delta\Omega_d$ is the recovery of spin frequency towards its pre-glitch value over time $\tau_d$. We have neglected the higher order derivative terms in the Taylor expansion and have also assumed single exponential recoveries. We then can construct a comparison matrix
\begin{table}[ht!]
\caption{Comparison matrix of various fit parameters of coupling between TD and the angular velocity such as $a_1$, $b_1$, $a_2$, $b_2$ and $c_2$, with reference to observed pulsar glitch characteristics such as $\phi$, $\Delta\Omega$ and $\tau$. The table summarizes the influence of different coupling strengths on glitch magnitude and energy dissipation.}
\label{comp}
\begin{center}
\begin{tabular}{ | m{5em} || m{1cm}| m{1cm} |m{1cm} |m{1cm} |m{1.5cm} | } 
  \hline
  \centering Fit Function & \centering $a_1$ & \centering $b_1$ & \centering $a_2$ & \centering $b_2$ & {\centering $c_2$} \\
  \hline
  \centering Glitch Model & \centering $\Delta\Omega_p$ & \centering $\Delta\phi_g$ &\centering $\Delta\Omega_d\tau_d$&\centering $\tau_d^{-1}$&$-\Delta\Omega_d\tau_d$ \\ 
  \hline
\end{tabular}
\end{center}
\end{table}
(table \ref{comp}) using which, various glitch data from pulsars can be compared to the defect model and corresponding $G\mu$ and parameter $B\Omega^2 l$ value can be estimated with the use of equation \ref{gmu}. Both of the parameters, $G\mu$ and $B\Omega^2 l$ are taken in such a way that they are dimensionless in the $G=c=1$ units. For the case of four pulsars discussed above we have listed the corresponding defect model values in (table \ref{def})
\begin{table*}[ht!]
\caption{The table lists observed or fictitious values for the change in angular velocity $\Delta\Omega$, decay time $\tau_d$, CS coupling $G\mu$, and dissipation parameter $B\Omega^2l$ for each pulsar. These values illustrate differences in glitch magnitude, decay behavior, and the strength of CS interactions across pulsars, including notable cases like the Vela and Crab pulsars.}
\label{def}
\begin{center}
\begin{tabular}{ | m{5 cm} || m{2cm}| m{3.5cm} |m{2cm} |m{2.5cm} |m{2.5cm} | } 
  \hline
  \hline
  \centering Name of Pulsar & $\Delta\Omega_p$ & $\tau_d$ & $G\mu$ $($G=c=1$)$ & $B\Omega^2 l$ $($G=c=1$)$ \\
  \centering  & ($10^{-5}$ Hz) & (d) & (1) & (1) \\
  \hline
  PSR J0835-4510 & 1.602 & 32 (Reported)& $2.128 \times 10^{-8}$ & $11655.6$ \\
  \hline
  PSR B0833-45 (Vela Pulsar)  & 1.596 & 0.0014 (Reported) &$2.121 \times 10^{-8}$ & $0.5103$\\
  \hline
  PSR B0531+21 (Crab Pulsar) & 1.530& 0.01 (Fictitious) & $2.032 \times 10^{-8}$& $3.6456$\\
  \hline
  PSR J1731-4744 & 0.379 & 0.5 (Fictitious) & $4.985 \times 10^{-9}$&$182.22$ \\
  \hline
\end{tabular}
\end{center}
\end{table*}
We have seen that the parameter $G\mu$ correlates with the glitch frequency of the pulsar whereas the parameter $B\Omega^2 l$ links with the glitch recovery time. It is seen that the Vela Pulsar, PSR J0835-4510 and the Crab pulsar has high angular velocity change which indicates significant glitch events on contrary to the PSR J1731-4744. The higher values of $G\mu$ for the former indicates a stronger coupling of the CS to the angular velocity 
 of the star during the PT than the latter which as a result gives rise to higher spin-up values for the former case. Also the decay time $\tau_d$ for the Vela and PSR J0835-4510 are reported observed data whereas, for Crab Pulsar and PSR J1731-4744, fictitious values are taken in this work for the sake of calculation. The corresponding parameter $B\Omega^2l$ value indicates how efficiently the gravitational wave energy from the CS cusps is dissipated from the NS, where lower value of the parameter indicates better dissipation than the higher ones. Thus if the gravitational wave energy is dissipated efficiently, then the pulsar's glitch relaxation can occur at a much faster rate.
 
\section{Continuous and burst Gravitational Wave Signals}
During the glitch event of a pulsar, for the corresponding $G\mu$ value, we have seen from the defect model that the release of gravitational energy can lead to the spin recovery phase of a glitch. This gravitational energy can be captured as a gravitational wave strain given by the equation \cite{PhysRevD.71.063510},
\begin{equation}
    h=C(f,t_0,z)\Gamma^{2/3}(G\mu)^{5/3}
\end{equation}
Where, $f$ is the frequency, $t_0$ is the time of occurrence and $z$ is the Redshift of the source. This gravitational wave emitted from the cusps of the CSs will be emitted along with the continuous emission of gravitational wave from the rotating NS. Thus the combined gravitational wave signal can be written as,
\begin{flalign}
    h_{comb}=& 
	h_0\sin\chi\left[\frac{1}{2}\cos\chi\sin{i}\cos{i}\cos\Omega \right. t  &&  \nonumber \\ 
	& \left. -\sin\chi\frac{1+\cos^2i}{2}\cos2\Omega t \right] \nonumber \\
        &+C(f,t_0,z)\Gamma^{2/3}(G\mu)^{5/3} \Theta(t-t_0)
	\label{wave}
\end{flalign}
\begin{figure}[ht!]
    \centering
    \includegraphics[scale=0.6]{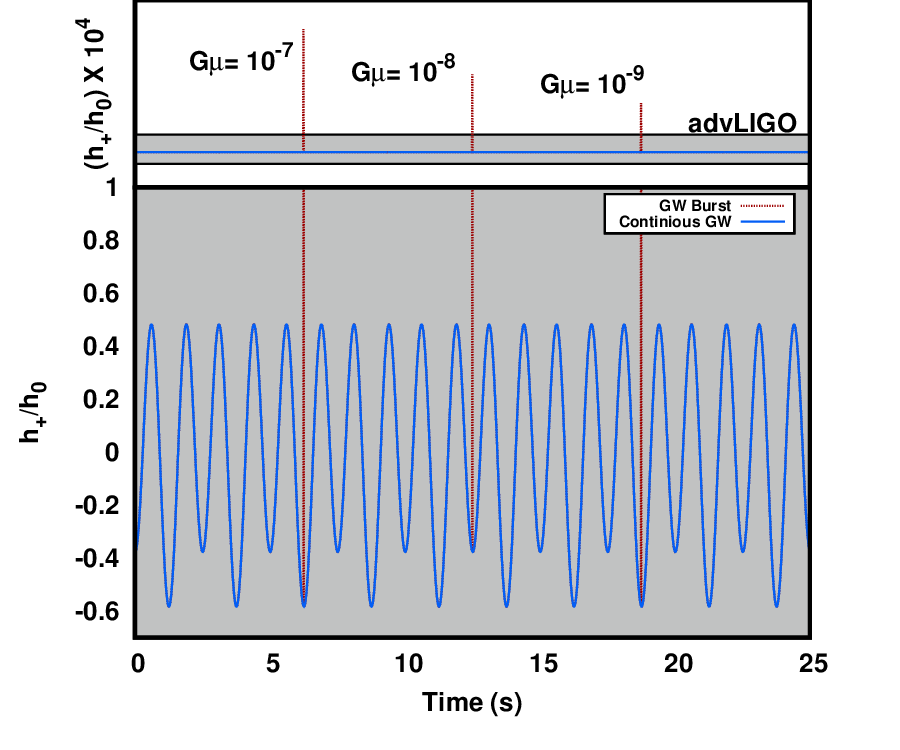}
    \caption{The plot presents a comparison between continuous gravitational waves (GW) and gravitational wave bursts, focusing on the strain amplitude $h_+/h_0$ as a function of time. The upper section shows sensitivity limits for gravitational wave detection, particularly from advLIGO, with varying values of $G\mu$. The blue waveform represents continuous gravitational waves (GW) over time due to rotating NS. The red vertical lines indicate the presence of gravitational wave bursts due to CS cusps at around $t=5$ s, $t=10$ s, and $t=20$ s (shown at different times only for comparison) for $G\mu=10^{-7}$, $10^{-8}$ and $10^{-9}$ respectively.}
    \label{fig4}
\end{figure}

The dependence of the strain signal on $(G\mu)^{5/3}$ implies that it depends heavily on the CS coupling and small change in the $G\mu$ value significantly effects strain profile. This has been show in (figure \ref{fig4}) where the continuous emission of gravitational waves is seen to also contain the burst signal from the coupled CS at three time instances $t_0$ (the Heaviside step function ensures that the burst part of the GW only appears at $t_0$ which is the time instance for the glitch event as well as the appearance of the TD) for $G\mu\approx 10^{-7}$, $10^{-8}$ and $10^{-9}$. The plot also contains the noise cut off for advLIGO which shows that the maximum strain for continuous emission is well under the noise curve of the detector however the burst signal remains above the noise curve which decreases as the $G\mu$ value decrease. The model indicates that the detectability of gravitational waves is influenced by the parameter $G\mu$. When $G\mu$ is higher, it enhances the strain, making it more likely for the gravitational wave signal to emerge from the background noise. 

The continuous emissions resulting from NS rotation might often be too faint to detect as they could fall below this threshold. However, a burst signal has the potential to briefly surpass this noise floor, particularly when  $G\mu$ is elevated.
Should future detectors enhance their sensitivity, even weaker bursts (with lower $ G\mu$) may become detectable, making interactions between CSs and NSs a promising source of gravitational wave signals.

\section{Conclusion}
In this study, we have explored the potential aspect of nontrivial TDs, particularly NG type CSs, to shed light on the observed rotational anomalies in NSs, including phenomena like pulsar glitches. NSs, characterized by their extreme densities and intense gravitational fields, provide a unique environment for examining the complex relationship between high-density physics and topological effects. Through our model of CSs' influence on the rotational dynamics of NSs, we investigated that these TDs can induce sudden spin-up events by modifying the internal coupling of angular velocity of the star with these defects formed during a PT event. This insight contributes a possible alternate approach to the understanding of glitches, against the existing models such as those based on superfluid behavior and crustal coupling.

Moreover, our study indicate that the interaction between CSs and NS rotation could result in the generation of distinctive gravitational wave signatures. These signatures may manifest as continuous emissions linked to the star’s rotational dynamics and burst-like emissions initiated by oscillations of the CSs cusps. Our analysis suggests that, under certain conditions, the gravitational wave strain associated with these events could be within the detection thresholds of current and upcoming observatories, such as advanced LIGO and Virgo. This could makes it plausible to empirically validate the existence of TDs in NSs. The relationship we identified between CS tension, angular velocity, and gravitational wave emissions is semi-universal, suggesting that the impact of TDs on NS dynamics may also apply across a wide range of NS models (EoS).

This study adds to the expanding knowledge base that links high-density astronomical phenomena with fundamental physics and highlights the potential role of TDs as one of the key players in observable astrophysical phenomena. Anticipated gravitational wave observations, together with improved pulsar timing arrays, could further illuminate these connections, leading to a more profound understanding of the intricate structures and extreme states of matter present in NSs. Our study however has various limitations such as, we have considered only NG type defects, however various other TD such as domain walls, monopoles, vortex lines, skyrmions etc could also significantly effect the rotation of the star in a similar (or may be different) manner. We also have not considered the further evolution of the CS once it is formed after PT apart from its cusps oscillation which could potentially explain multiple glitches of a single pulsar. We have also limited our study to glitch phenomena arising from puerly topological factors however in reality it could be a combined effect of such topological models and the crust coupling models. In our work we have limited our study to the above mentioned boundaries and the limitations mentioned forms our basis for future works.
\section{Acknowledgment}
DK would like to thank Raghunathpur College and the department of physics for providing the infrastructure for the research. DK would also like to thank Ritam Mallick and Rana Nandi for the insight into the equation of states.
\bibliographystyle{ieeetr} 
\bibliography{ref} 

\begin{thebibliography}{10}

\bibitem{Lasky_2015}
P.~D. Lasky, ``Gravitational waves from neutron stars: A review,'' {\em Publications of the Astronomical Society of Australia}, vol.~32, p.~e034, 2015.

\bibitem{annurev:/content/journals/10.1146/annurev-nucl-102419-124827}
J.~Lattimer, ``Neutron stars and the nuclear matter equation of state,'' {\em Annual Review of Nuclear and Particle Science}, vol.~71, no.~Volume 71, 2021, pp.~433--464, 2021.

\bibitem{annurev:/content/journals/10.1146/annurev-nucl-102711-094901}
H.-T. Janka, ``Explosion mechanisms of core-collapse supernovae,'' {\em Annual Review of Nuclear and Particle Science}, vol.~62, no.~Volume 62, 2012, pp.~407--451, 2012.

\bibitem{annurev:/content/journals/10.1146/annurev-astro-081915-023322}
F.~Özel and P.~Freire, ``Masses, radii, and the equation of state of neutron stars,'' {\em Annual Review of Astronomy and Astrophysics}, vol.~54, no.~Volume 54, 2016, pp.~401--440, 2016.

\bibitem{annurev:/content/journals/10.1146/annurev.aa.04.090166.002141}
J.~A. Wheeler, ``Superdense stars,'' {\em Annual Review of Astronomy and Astrophysics}, vol.~4, no.~Volume 4, 1966, pp.~393--432, 1966.

\bibitem{PhysRevC.107.025801}
R.~Somasundaram, I.~Tews, and J.~Margueron, ``Investigating signatures of phase transitions in neutron-star cores,'' {\em Phys. Rev. C}, vol.~107, p.~025801, Feb 2023.

\bibitem{doi:10.1142/S0218301318300084}
V.~Dexheimer, L.~T.~T. Soethe, J.~Roark, R.~O. Gomes, S.~O. Kepler, and S.~Schramm, ``Phase transitions in neutron stars,'' {\em International Journal of Modern Physics E}, vol.~27, no.~11, p.~1830008, 2018.

\bibitem{PhysRevD.107.083023}
J.~Zhu, C.~Wang, C.~Xia, E.~Zhou, and Y.~Ma, ``Probing phase transitions in neutron stars via the crust-core interfacial mode,'' {\em Phys. Rev. D}, vol.~107, p.~083023, Apr 2023.

\bibitem{Orsaria_2019}
M.~G. Orsaria, G.~Malfatti, M.~Mariani, I.~F. Ranea-Sandoval, F.~García, W.~M. Spinella, G.~A. Contrera, G.~Lugones, and F.~Weber, ``Phase transitions in neutron stars and their links to gravitational waves,'' {\em Journal of Physics G: Nuclear and Particle Physics}, vol.~46, p.~073002, jun 2019.

\bibitem{GLENDENNING2001393}
N.~K. Glendenning, ``Phase transitions and crystalline structures in neutron star cores,'' {\em Physics Reports}, vol.~342, no.~6, pp.~393--447, 2001.

\bibitem{sym16010111}
L.~Brandes and W.~Weise, ``Constraints on phase transitions in neutron star matter,'' {\em Symmetry}, vol.~16, no.~1, 2024.

\bibitem{MALLICK201496}
R.~Mallick and P.~Sahu, ``Phase transitions in neutron star and magnetars and their connection with high energetic bursts in astrophysics,'' {\em Nuclear Physics A}, vol.~921, pp.~96--113, 2014.

\bibitem{PODDER2024122796}
S.~Podder, S.~Pal, D.~Sen, and G.~Chaudhuri, ``Constraints on density dependent mit bag model parameters for quark and hybrid stars,'' {\em Nuclear Physics A}, vol.~1042, p.~122796, 2024.

\bibitem{Pan_2007}
N.-N. Pan and X.-P. Zheng, ``Observational constraints on quark matter in neutron stars,'' {\em Chinese Journal of Astronomy and Astrophysics}, vol.~7, p.~675, oct 2007.

\bibitem{Tangphati2022}
T.~Tangphati, I.~Karar, A.~Pradhan, and A.~Banerjee, ``Constraints on the maximum mass of quark star and the gw 190814 event,'' {\em The European Physical Journal C}, vol.~82, p.~57, Jan 2022.

\bibitem{PhysRevD.109.043054}
F.~M. da~Silva, A.~Issifu, L.~L. Lopes, L.~C.~N. Santos, and D.~P. Menezes, ``Bayesian study of quark models in view of recent astrophysical constraints,'' {\em Phys. Rev. D}, vol.~109, p.~043054, Feb 2024.

\bibitem{10.1093/mnras/stac3611}
S.~K. Maurya, K.~N. Singh, M.~Govender, and S.~Ray, ``{Observational constraints on maximum mass limit and physical properties of anisotropic strange star models by gravitational decoupling in Einstein–Gauss–Bonnet gravity},'' {\em Monthly Notices of the Royal Astronomical Society}, vol.~519, pp.~4303--4324, 12 2022.

\bibitem{condmat7010026}
A.~Maiellaro, F.~Illuminati, and R.~Citro, ``Topological phases of an interacting majorana benalcazar–bernevig–hughes model,'' {\em Condensed Matter}, vol.~7, no.~1, 2022.

\bibitem{PhysRevB.109.064503}
L.~M. Chinellato, C.~J. Gazza, A.~M. Lobos, and A.~A. Aligia, ``Topological phases of strongly interacting time-reversal invariant topological superconducting chains under a magnetic field,'' {\em Phys. Rev. B}, vol.~109, p.~064503, Feb 2024.

\bibitem{10.1093/nsr/nwt033}
S.-Q. Shen, ``{The family of topological phases in condensed matter†},'' {\em National Science Review}, vol.~1, pp.~49--59, 12 2013.

\bibitem{PhysRevB.99.125122}
D.~V. Else, H.~C. Po, and H.~Watanabe, ``Fragile topological phases in interacting systems,'' {\em Phys. Rev. B}, vol.~99, p.~125122, Mar 2019.

\bibitem{ZHITNITSKY1999647}
A.~R. Zhitnitsky, ``Topological defects and $\theta$ dependence in qcd.,'' {\em Nuclear Physics B - Proceedings Supplements}, vol.~73, no.~1, pp.~647--649, 1999.

\bibitem{doi:10.1142/9789812701725_0005}
A.~R. ZHITNITSKY, {\em TOPOLOGICAL DEFECTS IN QCD AT LARGE BARYON DENSITY AND/OR LARGE <inline-formula>N<sub>C</sub></inline-formula>}, pp.~60--71.

\bibitem{Higaki2016}
T.~Higaki, K.~S. Jeong, N.~Kitajima, T.~Sekiguchi, and F.~Takahashi, ``Topological defects and nano-hz gravitational waves in aligned axion models,'' {\em Journal of High Energy Physics}, vol.~2016, p.~44, Aug 2016.

\bibitem{Vachaspati:2015}
T.~Vachaspati, L.~Pogosian, and D.~A. Steer, ``{C}osmic strings,'' {\em Scholarpedia}, vol.~10, no.~2, p.~31682, 2015.
\newblock revision \#192547.

\bibitem{Blanco-Pillado_2023}
J.~J. Blanco-Pillado, D.~Jiménez-Aguilar, J.~Lizarraga, A.~Lopez-Eiguren, K.~D. Olum, A.~Urio, and J.~Urrestilla, ``Nambu-goto dynamics of field theory cosmic string loops,'' {\em Journal of Cosmology and Astroparticle Physics}, vol.~2023, p.~035, may 2023.

\bibitem{Sanyal2022}
S.~Sanyal, ``Nambu goto cosmic strings in the early universe,'' {\em The European Physical Journal Special Topics}, vol.~231, pp.~83--89, Jan 2022.

\bibitem{PhysRevD.91.083519}
A.~Lazanu, E.~P.~S. Shellard, and M.~Landriau, ``Cmb power spectrum of nambu-goto cosmic strings,'' {\em Phys. Rev. D}, vol.~91, p.~083519, Apr 2015.

\bibitem{zwiebach2009first}
B.~Zwiebach, {\em A First Course in String Theory}.
\newblock Cambridge University Press, 2009.

\bibitem{PhysRevLett.66.1126}
J.~R. Gott, ``Closed timelike curves produced by pairs of moving cosmic strings: Exact solutions,'' {\em Phys. Rev. Lett.}, vol.~66, pp.~1126--1129, Mar 1991.

\bibitem{WITTEN1985243}
E.~Witten, ``Cosmic superstrings,'' {\em Physics Letters B}, vol.~153, no.~4, pp.~243--246, 1985.

\bibitem{SARANGI2002185}
S.~Sarangi and S.-H. Tye, ``Cosmic string production towards the end of brane inflation,'' {\em Physics Letters B}, vol.~536, no.~3, pp.~185--192, 2002.

\bibitem{VILENKIN1985263}
A.~Vilenkin, ``Cosmic strings and domain walls,'' {\em Physics Reports}, vol.~121, no.~5, pp.~263--315, 1985.

\bibitem{Achúcarro2009}
A.~Ach{\'u}carro and C.~J. A.~P. Martins, {\em Cosmic Strings}, pp.~1641--1660.
\newblock New York, NY: Springer New York, 2009.

\bibitem{Hindmarsh_1995}
M.~B. Hindmarsh and T.~W.~B. Kibble, ``Cosmic strings,'' {\em Reports on Progress in Physics}, vol.~58, p.~477, may 1995.

\bibitem{Dave_2019}
S.~S. Dave and A.~M. Srivastava, ``Formation of topological vortices during superfluid transition in a rotating vessel,'' {\em Europhysics Letters}, vol.~126, p.~31001, may 2019.

\bibitem{PhysRevX.5.021015}
P.~M. Chesler, A.~M. Garc\'{\i}a-Garc\'{\i}a, and H.~Liu, ``Defect formation beyond kibble-zurek mechanism and holography,'' {\em Phys. Rev. X}, vol.~5, p.~021015, May 2015.

\bibitem{10.1093/mnras/stae1642}
P.~Bagchi, B.~Layek, D.~Saini, A.~Sarkar, A.~M. Srivastava, and D.~G. Venkata, ``{Detecting superfluid transition in the pulsar core},'' {\em Monthly Notices of the Royal Astronomical Society}, vol.~532, pp.~2934--2942, 07 2024.

\bibitem{Marmorini2024}
G.~Marmorini, S.~Yasui, and M.~Nitta, ``Pulsar glitches from quantum vortex networks,'' {\em Scientific Reports}, vol.~14, p.~7857, Apr 2024.

\bibitem{doi:10.1142/S0218271815300086}
B.~Haskell and A.~Melatos, ``Models of pulsar glitches,'' {\em International Journal of Modern Physics D}, vol.~24, no.~03, p.~1530008, 2015.

\bibitem{HASKELL2024102921}
B.~Haskell and D.~Jones, ``Glitching pulsars as gravitational wave sources,'' {\em Astroparticle Physics}, vol.~157, p.~102921, 2024.

\bibitem{Antonopoulou2022-vs}
D.~Antonopoulou, B.~Haskell, and C.~M. Espinoza, ``Pulsar glitches: observations and physical interpretation,'' {\em Rep Prog Phys}, vol.~85, Dec. 2022.

\bibitem{universe8120641}
S.~Zhou, E.~Gügercinoğlu, J.~Yuan, M.~Ge, and C.~Yu, ``Pulsar glitches: A review,'' {\em Universe}, vol.~8, no.~12, 2022.

\bibitem{PhysRevLett.116.169002}
Y.~V. Stadnik and V.~V. Flambaum, ``Stadnik and flambaum reply:,'' {\em Phys. Rev. Lett.}, vol.~116, p.~169002, Apr 2016.

\bibitem{Srivastava2017}
A.~M. Srivastava, P.~Bagchi, A.~Das, and B.~Layek, ``High-density qcd phase transitions inside neutron stars: Glitches and gravitational waves,'' {\em Pramana}, vol.~89, p.~68, Oct 2017.

\bibitem{10.1093/mnras/sty130}
M.~Antonelli, A.~Montoli, and P.~M. Pizzochero, ``{Effects of general relativity on glitch amplitudes and pulsar mass upper bounds},'' {\em Monthly Notices of the Royal Astronomical Society}, vol.~475, pp.~5403--5416, 01 2018.

\bibitem{Pizzochero_2011}
P.~M. Pizzochero, ``Angular momentum transfer in vela-like pulsar glitches,'' {\em The Astrophysical Journal Letters}, vol.~743, p.~L20, nov 2011.

\bibitem{Chang2022}
C.-F. Chang and Y.~Cui, ``Gravitational waves from global cosmic strings and cosmic archaeology,'' {\em Journal of High Energy Physics}, vol.~2022, p.~114, Mar 2022.

\bibitem{Gouttenoire2022}
Y.~Gouttenoire, {\em Gravitational Waves from Cosmic Strings}, pp.~419--499.
\newblock Cham: Springer International Publishing, 2022.

\bibitem{Auclair_2023}
P.~Auclair, S.~Blasi, V.~Brdar, and K.~Schmitz, ``Gravitational waves from current-carrying cosmic strings,'' {\em Journal of Cosmology and Astroparticle Physics}, vol.~2023, p.~009, apr 2023.

\bibitem{PhysRevD.110.063549}
K.~Schmitz and T.~Schr\"oder, ``Gravitational waves from low-scale cosmic strings,'' {\em Phys. Rev. D}, vol.~110, p.~063549, Sep 2024.

\bibitem{Sousa2024}
L.~Sousa, ``Cosmic strings and gravitational waves,'' {\em General Relativity and Gravitation}, vol.~56, p.~105, Sep 2024.

\bibitem{PhysRevLett.85.3761}
T.~Damour and A.~Vilenkin, ``Gravitational wave bursts from cosmic strings,'' {\em Phys. Rev. Lett.}, vol.~85, pp.~3761--3764, Oct 2000.

\bibitem{PhysRevC.76.045801}
S.~K. Dhiman, R.~Kumar, and B.~K. Agrawal, ``Nonrotating and rotating neutron stars in the extended field theoretical model,'' {\em Phys. Rev. C}, vol.~76, p.~045801, Oct 2007.

\bibitem{PhysRevC.66.064302}
T.~Nik\ifmmode \check{s}\else \v{s}\fi{}i\ifmmode~\acute{c}\else \'{c}\fi{}, D.~Vretenar, and P.~Ring, ``Relativistic random-phase approximation with density-dependent meson-nucleon couplings,'' {\em Phys. Rev. C}, vol.~66, p.~064302, Dec 2002.

\bibitem{PhysRevC.97.045806}
B.~Kumar, S.~K. Patra, and B.~K. Agrawal, ``New relativistic effective interaction for finite nuclei, infinite nuclear matter, and neutron stars,'' {\em Phys. Rev. C}, vol.~97, p.~045806, Apr 2018.

\bibitem{PhysRevC.66.055803}
C.~J. Horowitz and J.~Piekarewicz, ``Constraining urca cooling of neutron stars from the neutron radius of ${}^{208}\mathrm{Pb}$,'' {\em Phys. Rev. C}, vol.~66, p.~055803, Nov 2002.

\bibitem{1971ApJ...170..299B}
G.~{Baym}, C.~{Pethick}, and P.~{Sutherland}, ``{The Ground State of Matter at High Densities: Equation of State and Stellar Models},'' {\em \apj}, vol.~170, p.~299, Dec. 1971.

\bibitem{Romani_2022}
R.~W. Romani, D.~Kandel, A.~V. Filippenko, T.~G. Brink, and W.~Zheng, ``Psr j0952-0607: The fastest and heaviest known galactic neutron star,'' {\em The Astrophysical Journal Letters}, vol.~934, p.~L17, jul 2022.

\bibitem{PhysRevD.109.043052}
Y.-Z. Fan, M.-Z. Han, J.-L. Jiang, D.-S. Shao, and S.-P. Tang, ``Maximum gravitational mass ${M}_{\mathrm{tov}}=2.2{5}_{\ensuremath{-}0.07}^{+0.08}{M}_{\ensuremath{\bigodot}}$ inferred at about 3

\bibitem{Riley_2021}
T.~E. Riley~et al, ``A nicer view of the massive pulsar psr j0740+6620 informed by radio timing and xmm-newton spectroscopy,'' {\em The Astrophysical Journal Letters}, vol.~918, p.~L27, sep 2021.

\bibitem{Miller_2021}
M.~C. Miller~et al, ``The radius of psr j0740+6620 from nicer and xmm-newton data,'' {\em The Astrophysical Journal Letters}, vol.~918, p.~L28, sep 2021.

\bibitem{Abbott_2020}
R.~Abbott~et al and {LIGO Scientific Collaboration and Virgo Collaboration}, ``Gw190814: Gravitational waves from the coalescence of a 23 solar mass black hole with a 2.6 solar mass compact object,'' {\em The Astrophysical Journal Letters}, vol.~896, p.~L44, jun 2020.

\bibitem{Levon_2009}
L.~Pogosian, S.-H.~H. Tye, I.~Wasserman, and M.~Wyman, ``Cosmic strings as the source of small-scale microwave background anisotropy,'' {\em Journal of Cosmology and Astroparticle Physics}, vol.~2009, p.~013, feb 2009.

\bibitem{Jeong_2005}
E.~Jeong and G.~F. Smoot, ``Search for cosmic strings in cosmic microwave background anisotropies,'' {\em The Astrophysical Journal}, vol.~624, p.~21, may 2005.

\bibitem{1967ApJ...150.1005H}
J.~B. {Hartle}, ``{Slowly Rotating Relativistic Stars. I. Equations of Structure},'' {\em \apj}, vol.~150, p.~1005, Dec. 1967.

\bibitem{PhysRevD.91.063007}
O.~Hamil, J.~R. Stone, M.~Urbanec, and G.~Urbancov\'a, ``Braking index of isolated pulsars,'' {\em Phys. Rev. D}, vol.~91, p.~063007, Mar 2015.

\bibitem{PhysRevD.71.063510}
T.~Damour and A.~Vilenkin, ``Gravitational radiation from cosmic (super)strings: Bursts, stochastic background, and observational windows,'' {\em Phys. Rev. D}, vol.~71, p.~063510, Mar 2005.

\bibitem{10.1093/mnras/stz3230}
A.~Basu, B.~C. Joshi, M.~A. Krishnakumar, D.~Bhattacharya, R.~Nandi, D.~Bandhopadhay, P.~Char, and P.~K. Manoharan, ``Observed glitches in eight young pulsars,'' {\em Monthly Notices of the Royal Astronomical Society}, vol.~491, pp.~3182--3191, 11 2019.

\bibitem{Ashton2019}
G.~Ashton, P.~D. Lasky, V.~Graber, and J.~Palfreyman, ``Rotational evolution of the vela pulsar during the 2016 glitch,'' {\em Nature Astronomy}, vol.~3, pp.~1143--1148, Dec 2019.

\bibitem{10.1093/mnras/sty1294}
B.~Shaw, A.~G. Lyne, B.~W. Stappers, P.~Weltevrede, C.~G. Bassa, A.~Y. Lien, M.~B. Mickaliger, R.~P. Breton, C.~A. Jordan, M.~J. Keith, and H.~A. Krimm, ``The largest glitch observed in the crab pulsar,'' {\em Monthly Notices of the Royal Astronomical Society}, vol.~478, pp.~3832--3840, 05 2018.

\end{thebibliography}

\end{document}